\documentclass[aps,prb,reprint,superscriptaddress,amsmath,amssymb,floatfix,nofootinbib]{revtex4-2}
\usepackage{chngcntr}
\usepackage{graphicx}
\usepackage[svgnames]{xcolor}
\usepackage[colorlinks=true,linktoc=page,linkcolor=blue,citecolor=blue,urlcolor=blue]{hyperref}
\usepackage[capitalise]{cleveref}
\usepackage{physics,mathtools,mathrsfs}
\usepackage{bm}

\usepackage{microtype} 
\usepackage[english]{babel}

\usepackage{mathrsfs}
\usepackage{slashed}
\usepackage{comment}

\def\v[#1]{\textbf{#1}}
\def\w[#1]{\widehat{#1}}
\def\vs[#1,#2]{\boldsymbol{{#1}_{#2}}}
\def\mes[#1]{d^{3}{#1}}

\def\<{\langle}
\def\>{\rangle}
\def\vecs[#1,#2]{\boldsymbol{{#1}_{#2}}}

\newcommand{\be}{\begin{equation}}
	\newcommand{\ee}{\end{equation}}
\newcommand{\bes}{\begin{subequations}}
	\newcommand{\ees}{\end{subequations}}

\def\e{\epsilon}

\def\nn{\nonumber}

\begin{document}
	
	\title{Emergent Carroll symmetry at phase separation in one-dimensional lattice systems}
	\author{Sourav Biswas}
	\email{sourav.biswas@dipc.org}
	\affiliation{Indian Institute of Technology Kanpur, Kanpur 208016, India}
	\affiliation{DIPC - Donostia International Physics Center, Paseo Manuel de Lardiz{\'a}bal 4, 20018 San Sebasti{\'a}n, Spain}
	\author{Ashutosh Dubey}
	\email{adubey@iitk.ac.in}
	\author{Saikat Mondal}
	\email{saikatmd@iitk.ac.in}
	\affiliation{Indian Institute of Technology Kanpur, Kanpur 208016, India}
	\author{Aritra Banerjee}
	\email{aritra.banerjee@pilani.bits-pilani.ac.in}
	\affiliation{Birla Institute of Technology and Science, Pilani Campus, Pilani, Rajasthan 333031, India.}
	\author{Arijit Kundu}
	\email{kundua@iitk.ac.in }
	\affiliation{Indian Institute of Technology Kanpur, Kanpur 208016, India}
	\author{Arjun Bagchi.} 
	\email{abagchi@iitk.ac.in}
	\affiliation{Indian Institute of Technology Kanpur, Kanpur 208016, India}

	\begin{abstract}
		Asymptotic behavior of generic Tomonaga-Luttinger liquid in the vicinity of phase-separated regions is known to produce an instability where well-known relativistic Conformal Field Theory (CFT) techniques fail. In this paper, we introduce an analytic paradigm that provides a continuum description of this important issue. We show that there is an emergent Carrollian symmetry when phase separation is reached, and techniques of Carroll CFT, as opposed to its relativistic relative, are central to the understanding of the physics. We work with the analogous spinless fermionic system in this region and capture the transition across this phase separation. Our numerical results corroborate the density-density correlations intrinsically computed using Carroll CFT. We further test the framework in a number of lattice systems, namely the spinless and spinfull fermionic models with distinct microscopic content, and find the same scaling at the transition. We discuss the scope of the framework and broader perspective.
	\end{abstract}
	
	\maketitle
	
	\section{Introduction}\label{sec:intro}
	The theory of phase transitions is intimately connected with conformal symmetry. Critical phenomena and their universality in the modern day is studied through relativistic conformal field theories (CFT) \cite{DiFrancesco:1997nk}, since at critical points all scales become equally important and there is emergent scaling symmetry. However, it is far from obvious why scaling has to come hand-in-hand with an underlying Poincaré symmetry and hence why relativistic CFTs would be all encompassing in the study of critical phenomena, especially since there exist many condensed matter systems without underlying relativistic structure. In this paper, we find an explicit example of a class of systems where there is a region in parameter space governed by a scale invariant theory, but with relativistic symmetry replaced by \textit{Carroll} symmetry. 
	
	Carroll symmetry is an ultra-relativistic version of the better known Galilean symmetry. The Galilean algebra can be obtained by sending the speed of light to infinity $(c\to\infty)$ in the Poincaré algebra. To obtain the Carroll algebra, $c$ is instead dialled to zero \cite{SenGupta:1966qer, LBLL}, making the system manifestly \textit{ultralocal}.  
	
	A broad class of phenomena in many-body condensed matter physics is governed by the principles of relativistic CFTs \cite{DiFrancesco:1997nk, GiamarchiB, FradkinB, Voit1995, Senechal2004, NagaosaQFT, Affleck:1990iv, Affleck:1990zd, GiaQuenCft} and the (Tomonaga-) Luttinger Liquid (LL) \cite{GiamarchiB, FradkinB, Voit1995, Senechal2004, NagaosaQFT, Mattlieb, Lutt63} is one of the most prominently studied CFTs in this context. LL categorizes an extensive range of many body phases in one dimensional (1D) and quasi-1D systems \cite{QSHEdge, Kane, Glazman, Wen, Haldane1981, FuruKondo, SB1, SB2, SB3}. Well known relativistic CFT techniques are known to fail \cite{japaninstability,GiamarchiB,Franchini2017,Cabra2004} at a particular point in parameter space of such LL-like theories,  which is referred to as \textit{Phase Separation} (PS) \cite{japaninstability, Lewen2021, Moreno, Batrouni}. Analytic studies of the PS region have thus been fraught with difficulties. In this paper, we put forward a {continuum-level description} of this problem and suggest that a new paradigm is required to tackle this issue that is applicable to all systems admitting a LL-like description. 
	
	We show that the approach to PS region is directly related to the emergence of Carrollian symmetry. Physics at PS is naturally captured by Carrollian CFTs, much like in many systems relativistic CFTs dictate behaviour at critical points. Our central message is: the physics of phase separation in the Luttinger Liquid parameter space, where conventional methods fail, is naturally described by an underlying emergent Carrollian description that helps to gain analytic control over the region. 
	
	We organize our claims along two distinct lines. At the continuum level, the bosonized Luttinger Hamiltonian reduces in the $u \to 0$ limit to that of an electric Carrollian scalar field theory, whose symmetry generators and Ward identities have been worked out in detail in the recent literature \cite{Henneaux:2021yzg, Bagchi:2022emh, deBoer:2023fnj, Banerjee:2023jpi}. We use those results to predict a quadratic small-$q$ dependence of the static density--density structure factor and an extensive low-energy manifold associated with supertranslations. At the lattice level, we test these predictions in the $t$--$V$ model of spinless fermions and in the fermionic $tJ$ model using density-matrix renormalization group calculations. The numerical observations, the crossover of $S(q) \sim |q|$ to $S(q) \sim q^{2}$, the divergent compressibility, the particle-hole segregation, and the extensive degeneracy, are direct outputs of the analysis. The identification of the PS boundary with an emergent Carrollian CFT, with the consequent interpretive content, follows from matching these features with the continuum predictions. We are explicit about this distinction in our concluding discussion.
	
	This work highlights the role of Carroll symmetries in phase transitions and adds to the growing body of work which studies the relevance of Carroll physics in condensed matter systems \cite{Bidussi:2021nmp, Marsot:2022imf, Bagchi:2022eui,  Figueroa-OFarrill:2023vbj,Ara:2024fbr}.

	\section{Luttinger liquid}\label{sec:LL}
	The LL, in the continuum description, is an effective massless scalar field theory corresponding to the long-wavelength limit of an underlying interacting lattice model. It describes the density fluctuations of a 1D interacting fermion system and can be written as a bosonic CFT with unit central charge. The Hamiltonian of the system is \cite{GiamarchiB, FradkinB, Voit1995, Senechal2004}
	\begin{eqnarray}\label{masterH}
		\mathcal{H}  = \frac{1}{2 \pi} \int dx \Big[ uK  (\pi \Pi(x))^2 +  \frac{u}{K}  (\partial_x \phi(x))^2\Big],
	\end{eqnarray}
	where $u$ and $K$ are the effective velocity and Luttinger parameter respectively. The system is modelled in terms of dual bosonic fields $\phi(x)$ and $\theta(x)$ such that the conjugate momentum to $\phi(x)$ is $\Pi(x) = \frac{1}{\pi}\partial_x\theta(x)$. In the continuum description, this maps to a bosonic field compactified on a circle. The above Hamiltonian also results from Bosonization that combines a free theory of spinless fermions with a density-density interaction: $ \mathcal{H} =  \mathcal{H}_0 +  \mathcal{H}_{int}$ where, $\mathcal{H}_0$ is the field-theoretic equivalent of Nearest-Neighbour (NN) hopping
	\begin{eqnarray}
		\mathcal{H}_0 = -i v_F \int dx ~\Psi^{\dagger}(x) \partial_x \Psi(x).
	\end{eqnarray}
	$v_F$ is the Fermi velocity of the free theory. The standard density-density interaction is 
	\begin{eqnarray}
		\mathcal{H}_{int} = \int\,dx dx' \mathcal{V}(x-x')\rho(x)\rho(x'),
	\end{eqnarray}
	where $ \mathcal{V}(x-x')$ is the lattice model dependent effective NN interaction in position space. We define local densities of left and right-moving fermions as $\rho_r(x) = \Psi_r^\dagger(x) \Psi_r(x)\,,r=R,L=\pm$, whereas the total density is $\rho(x)=\rho_R(x)+\rho_L(x)$. Fermionic field operators are related to the bosonic fields via bosonization as $\Psi_{r}(x) = \frac{1}{\sqrt{2\pi \epsilon}}e^{-i (r\phi(x) - \theta(x))}$. 
	
	The full Hamiltonian, diagonalised by a Bogoliubov transformation, takes the form of \eqref{masterH} with effective Luttinger parameters $u$ and $K$ 
	\footnote{With $u$ and $K$ being determined by coupling constants of the interaction terms and $v_F$. }. In this picture, the PS region is a special point achieved by taking $u \to 0$, such that $uK \to 1$ but $u/K \to 0$.
	Such a point is ill-defined in the LL paradigm and understood as an instability to LL \cite{japaninstability, GiamarchiB}.

	\section{Lattice System}\label{sec:lattice}
	We now focus on a specific lattice Hamiltonian known as the $t-V$ model \cite{GiamarchiB, japaninstability} describing a system of spinless fermions ($c^{\dagger}_j,c_j$) on a 1D lattice, with NN interactions. The Hamiltonian is written as
	\begin{align}\label{H_tV}
		\mathcal{H}_{t-V} &= -t \sum_i ( c^{\dagger}_{i} c_{i+1} + c^{\dagger}_{i+1} c_{i} ) - V \sum_i n_i n_{i+1},
	\end{align}	
	where $n_j = c^{\dagger}_j c_j$ is the fermion number operator (with  $\{c_i,c_j^{\dagger}\} = \delta_{ij}$) and $V$ is interaction strength. We set $t=1$. 
	
	At the long wavelength (momenta $q \to 0$) limit, the lattice model can be bosonized to connect with the field theoretic Hamiltonian \eqref{masterH}. Bosonization gives an effective description for the ideal LL regime ($|V|<2$). Our study probes beyond this limit. The LL picture is not valid anymore in this regime, however numerical results continue to help us navigate. The parameters $u, K$ can be expressed as $u = 2\sqrt{1 -(V/2)^2} $ and $ K = \sqrt{\frac{ 1+V/2}{1-V/2} }$, \footnote{Terms like $\rho_\nu \rho_\nu$ (with $\nu =R,L$) do not create relative differences between $uK$ and $u/K$, apart from an overall shift in energy and we drop them. Here, $V$ is assumed to be scaled by $\pi$. At half-filling $v_F= 2.$} \cite{GiamarchiB, FradkinB}. The PS region is at $V\to 2$, with $u \to 0, K \to \infty$  \footnote{Another well-known example of a 1D lattice model in the LL class is the spin-1/2 quantum XXZ Heisenberg chain that can be exactly mapped to \eqref{masterH} at half-filling. This is a famous Bethe-ansatz integrable model of nearest-neighbor coupled spins, except for when the anistropy $\Delta =1$, where the phase separation happens. A ferromagnetic phase prevails at $\Delta \geq 1$. }.
	
	From \eqref{masterH}, we see exactly at $V=2$ the Hamiltonian is just $\sim \Pi(x)^2$. This region, previously thought of as inaccessible, is precisely the Carroll region, as we explain below.

	\section{Conformal Carroll symmetry}\label{sec:CarrollSym}
	An Inönü-Wigner contraction from the Poincaré group taking $c\to0$ gives rise to the Carroll group \cite{SenGupta:1966qer, LBLL} (detailed algebra is provided in the appendix A of the Supplementary Material \cite{Suppl}). Commutativity of boosts distinguishes the Carroll group from its relativistic parent.  The Conformal Carroll group additionally contains dilatations, and Carroll special conformal transformations. Carroll symmetries can emerge as a characteristic velocity scale, e.g. the (effective) Fermi velocity, in a physical system goes to zero. A notable example is twisted bilayer graphene \cite{Bagchi:2022eui}, where tuning to ``magic" angle corresponds to this limit. Carroll symmetries are of great recent relevance arising in holographic duals of asymptotically flat spacetimes \cite{Bagchi:2010zz,Bagchi:2012cy,Bagchi:2012xr,Barnich:2012xq,Bagchi:2016bcd,Donnay:2022aba,Bagchi:2022emh,Donnay:2022wvx,Bagchi:2023fbj}, on black hole horizons \cite{Donnay:2019jiz}, in cosmology \cite{deBoer:2021jej} and string theory \cite{tens1,tens2,tens3,tens4}. The full conformal Carrollian algebra, its representation theory, and the corresponding Ward identities and conserved currents for the electric scalar theory used in this paper have been worked out in detail in \cite{Bagchi:2022emh,Henneaux:2021yzg,deBoer:2023fnj,Banerjee:2023jpi}; we adopt those results without rederivation.
	
	In this paper, we will compute correlation functions. Here the change from relativistic to Carroll symmetry shows up distinctively. The Carroll CFTs two-point function, fixed by Ward identities, is very different from relativistic CFTs. For Carroll primary fields $(\Phi_\Delta)$, this is \cite{Bagchi:2022emh} \footnote{There is also a time-independent solution to the Ward identities which we will not be interested in.}
	\begin{equation}\label{deltabranch}
		\langle \Phi_\Delta(t,\vec{x}) \Phi_{\Delta'}(t',\vec{x'})\rangle \sim \frac{\delta^{(d)}(\vec{x}-\vec{x'})}{(t-t')^{\Delta + \Delta'-d}}.
	\end{equation}
	$\Delta, \Delta'$ are conformal dimensions of the fields (more details in the appendix).
	
	\section{Carroll scalar field theory}\label{sec:Cscalar}
	Carroll field theories can be classified into  \textit{Electric} and  \textit{Magnetic} theories \cite{deBoer:2021jej, Henneaux:2021yzg}. In Electric theories, temporal derivatives dominate and this connects directly with the PS region in the continuum Luttinger Hamiltonian \eqref{masterH}. The Electric Carroll  scalar action is:
	\be{}
	\mathcal{S} = \frac{1}{2}\int dt d^dx \bigg((\partial_t \phi)^2 - m^2\phi^2 \bigg).
	\ee
	The equation of motion reads  $(\partial_t^2 + m^2) \phi(t,\vec{x}) = 0$. The most general solution is \cite{Banerjee:2023jpi},
	\begin{eqnarray}
		\phi(t,\vec{x}) = \frac{1}{\sqrt{m}} \bigg( a^\dagger (\Vec{x})e^{imt} + a(\Vec{x})e^{-imt}\bigg)
	\end{eqnarray}
	where $a, a^\dagger$ are annihilation/creation operators obeying $[a(\Vec{x}) , a^\dagger (\Vec{x'})] = \frac{1}{2}\delta^{(d)}(\vec{x} - \vec{x'})$. The Carroll Hamiltonian is
	\begin{eqnarray}
		H = m\int d^dx\, \bigg( 2a^\dagger (\Vec{x})a(\Vec{x}) + \frac{1}{2}\delta^{(d)}(0)\bigg).
	\end{eqnarray}
	The zero-point energy $\delta^{(d)}(0)$ can be eliminated by normal ordering. 
	We define the ground state $|0\rangle$, satisfying $a(\Vec{x})|0\rangle = 0, \, \forall \vec{x}$. The first excited state $a^\dagger (\Vec{x})|0\rangle$ has energy $m$ for any $\vec{x}$ and thus comes with infinite degeneracy. These states are not momentum eigenstates, and defining translation symmetric or Bloch-like states  requires care \cite{Ara:2024fbr}. This is a reflection of the inherent infinite \textit{supertranslation} symmetry of Carroll theories \footnote{In the continuum theory these are generated by the vector fields $M_f = f(x) \partial_t$ with $f(x)$ being an arbitrary function. For a discussion on discrete supertranslation on a lattice, see \cite{Ara:2024fbr}.} that leads to Hamiltonians with trivial dispersion \cite{Henneaux:2021yzg, Bagchi:2022eui}. 
	
	After defining $|0\rangle$, the time-ordered two-point function for massive Carroll scalars become:
	\begin{equation}\label{car-del}
		\langle \mathcal{T} (\phi(t,\vec{x}) \phi(t',\vec{x'}))\rangle = \frac{1}{2m}e^{-im|t-t'|}\delta^{(d)}(\vec{x}-\vec{x'}).
	\end{equation}
	We see the naive $ m\rightarrow 0$ limit leads to divergent results. Appropriately regulating, we get  
	\begin{equation}\label{deltam0}
		\langle \mathcal{T} (\phi(t,\vec{x}) \phi(t',\vec{x'}))\rangle = -\frac{i}{2}|t-t'|\delta^{(d)}(\vec{x}-\vec{x'}).
	\end{equation}
	The spatial delta-functions in \eqref{car-del}, \eqref{deltam0} also appear in higher point scalar correlators and in correlations of other (Electric) Carroll theories, reflecting their ultralocal nature. The explicit computation for the scalar theory matches with \eqref{deltabranch} found from symmetry arguments since $\Delta = (d-1)/2$. For some more details of two-point correlation functions, we point the reader to the Supplementary Material \cite{Suppl}.

	\section{Density-Density correlation}\label{sec:rhorho}
	It is clear from the Hamiltonians \eqref{masterH} and \eqref{H_tV} that the physics of PS region should be directly related to the appearance of Carroll structure. We now focus on physical quantities computed from Carroll theories and make predictions of the PS region, which we later reproduce by numerical results. 
	
	We examine the density-density correlator as the physics, dictated by scalar fields, moves from a relativistic to a Carroll regime. In a fermionic system, the density operator can be thought of as the fluctuation from average density of particle $\rho_0$. Where bosonization techniques are allowed, the counterpart field $\phi$ is used to write down this fluctuation \cite{GiamarchiB}:
	\begin{eqnarray}
		\rho(x)= \bigg(\rho_0 - \partial_x\phi(x)\bigg)\sum\limits_{p\in \mathbb{Z}}e^{2\pi ip(\rho_0x - \phi(x))}.
	\end{eqnarray}
	For a low-energy effective theory, in general, only the slowly varying modes around $p=0$ are  picked up and the density fluctuations can be written \cite{GiamarchiB, FradkinB, Voit1995} as $\delta\rho(x)\approx \partial_x\phi$. Away from the PS region computation of density correlators require the two point correlation between conformal scalar fields given by $\langle \phi(x)\phi(x') \rangle = -\frac{1}{2K}\log(|x-x'|)$.
	Hence, the density-density correlator is
	\begin{eqnarray}
		\langle \rho(x) \rho(x')\rangle &&= \langle \partial_x\phi(x) \partial_x\phi(x')\rangle - \langle \partial_x\phi(x) \rangle \langle \partial_x\phi(x') \rangle, \nonumber\\ 
		&&= \frac{1}{2K}\frac{1}{|x-x'|^2}.
	\end{eqnarray}
	In momentum space:
	\begin{equation}\label{Eq:Br1q}
		\langle \rho(q) \rho(-q)\rangle = -\sqrt{\frac{\pi}{2}}\frac{q}{2K}\text{sign}(q).
	\end{equation}
	This is the well-known linear variation of density-density correlator as a function of $q$ in the small momentum regime, which is true also for LL, as here the relativistic CFT map works perfectly.
	
	Now as we hit PS, the electric Carroll region takes over. Two-point function of relevance is \eqref{deltam0} with $d=1$. The density-density correlator in the Carroll regime is thus \footnote{There is a $(t-t')$ term in the correlator as well which we appropriately regulate when taking the equal time limit.}
	\begin{equation}
		\langle \rho(x) \rho(x')\rangle \sim \partial_x^2 \delta(x - x'),
	\end{equation}
	which in momentum space simply reads
	\begin{equation}\label{Eq:Br2q}
		\langle \rho(q) \rho(-q)\rangle \sim -\frac{1}{\sqrt{2\pi}}q^2.
	\end{equation}
	So we expect density-density correlations calculated in momentum space to move from a linear to a quadratic dependence on momenta as one hits the PS from the LL phase. This would be the transition structure expected in a system that undergoes a change from a relativistic to a Carroll CFT. We note that an $S(q)\sim q^{2}$ form is, taken on its own, also consistent with the more conventional viewpoint based on the vanishing of the sound velocity and the divergence of the compressibility at PS. What the Carrollian framework adds, beyond reproducing this scaling, is a single structural origin for the simultaneous occurrence of the $q^{2}$ form, the divergent compressibility, and the extensive ground-state degeneracy associated with supertranslations; we examine the latter two features in the lattice numerics below. 
    We also would like to point out here that this more conventional picture of vanishing of sound velocity is directly related to the emergence of Carrollian symmetry, since as stressed in the introduction, the vanishing of any characteristic velocity (in this case, the velocity of sound) is tantamount to the closing of an analogue ``light'' cone.

	\section{Numerical results}\label{sec:numerics}
	We now discuss an intrinsic numerical way to investigate the PS region for the LL theory. We show that the results from many-body numerics in the PS parameter space are consistent with the predictions from Carroll CFT. We employ DMRG \cite{White,SCHOLL} for numerical analysis of the interacting lattice model \eqref{H_tV} with the help of ITensor library \cite{itensor}. We use a 1D lattice with $L$ sites, the maximum allowed bond dimension $\chi_{dim}$ is varied between $10^{3}$ to $10^{4}$ as per convergence requirement, and the truncation error for the ground state energy is maintained at the order of $ 10^{-10}$. We note how the scaling of density-density correlation in Eqs.~\eqref{Eq:Br1q},\eqref{Eq:Br2q} encapsulates the physics of two separate regimes of scale invariant physics discussed earlier. In case of the lattice system, the properties of aforementioned density-density correlation is captured instead by the density-density structure factor in momentum space:
	\begin{equation}
		\begin{aligned}\label{Sq}
			S (q) &&= \frac{1}{L} \sum_{i,j} \big[ \langle n_i n_j \rangle -\langle n_i \rangle \langle n_j \rangle \big] e^{iq(i-j)}.
		\end{aligned}	
	\end{equation}
	Here $n_i$ and $n_j$ are particle densities at different sites. 
	The numerical evaluation of $S (q)$ efficiently shows the transition across $V=2$ through the different scaling behaviors. The exponent of scaling for $S(q)$ is extracted numerically for $V<2$ and $V>2$. One can extract this scaling exponent as a function of $q$, written as:
	\begin{equation}
		\begin{aligned}\label{epsilon}
			\epsilon(q) = \frac{d \log S(q)}{d \log q} .
		\end{aligned}	
	\end{equation}

	\begin{figure}
		\centering
		\includegraphics[width=.45\textwidth]{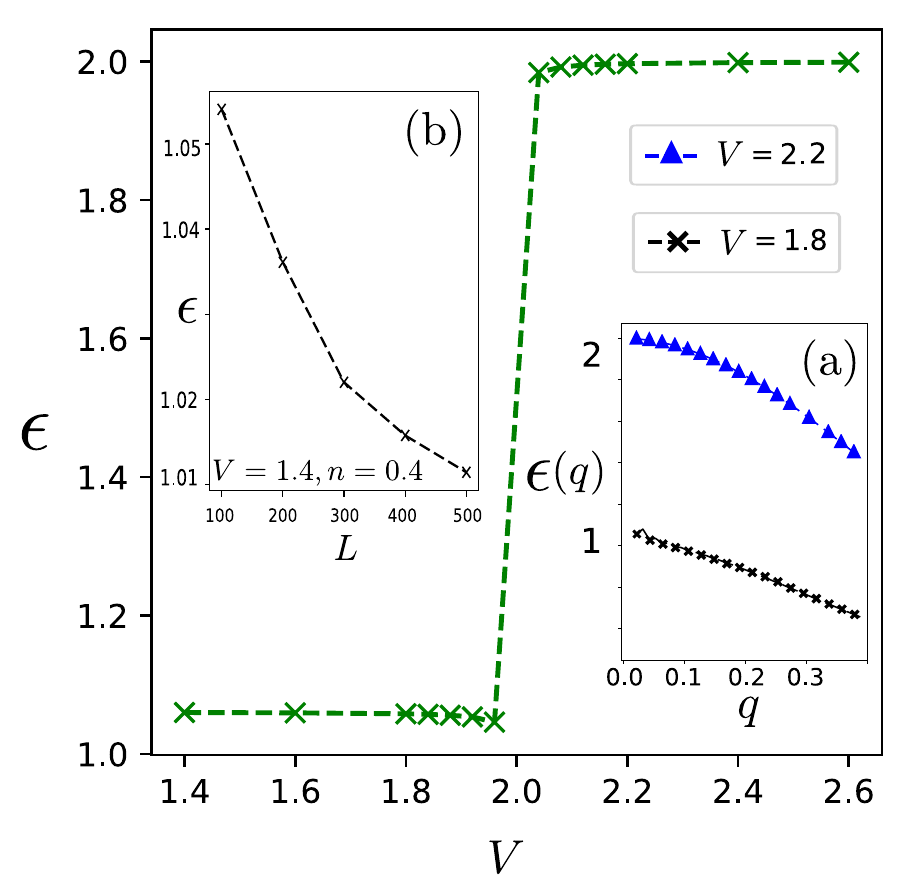}
		\caption{ Plot of the scaling exponent ($\epsilon$) as a function of $V$, for $L=300$. As we scan across the $V$-axis, the transition becomes evident around $V=2$. We show result for $n=0.5$, but the same can be done for other filling factors shown in Fig.~\eqref{fig:Compr}. In the inset (a), we show $\epsilon(q)$ vs $q$ for different $V$s. We show one instance of finite size scaling (at $V=1.4, n=0.4$) to strengthen our results, in inset (b). The exponent approaches 1 with increasing system size.}\label{fig:ExDer}
	\end{figure}

	We denote the $q \to 0$ limit of this quantity as $\epsilon = \epsilon(q) |_{q \to 0}$. We refer to Fig.(\ref{fig:ExDer}) for a demonstration of the results. For, $V<2$ we get $\epsilon \sim 1$, implying a $c=1$ CFT with linear density correlation. For, $V>2$ the ground state has domain walls, separating particle and hole like regions. This phase jumps to a $\epsilon \sim 2$ scaling across the $V=2$ point. This is what we had predicted from our analytical Carroll model. 
	
	From the limit to get to PS, it is obvious looking at the Hamiltonian that the system has an emergent Carroll symmetry at that exact point. We have now demonstrated from an independent numerical analysis that the density-density correlations lead to $\epsilon=2$, matching the prediction of the Carroll model. 
	
	We now provide further evidence for this emergent symmetry. As mentioned in passing above \eqref{car-del}, our Carroll model has an infinite dimensional {\em supertranslation} invariance. This leads to an infinite vacuum degeneracy of the continuum model. In the lattice model probed by our numerics, we see the emergence of this vacuum degeneracy. The interested reader is pointed to the Supplementary Material \cite{Suppl} for more details on this. 
	
\begin{figure}
		\centering
		\includegraphics[width=.485\textwidth]{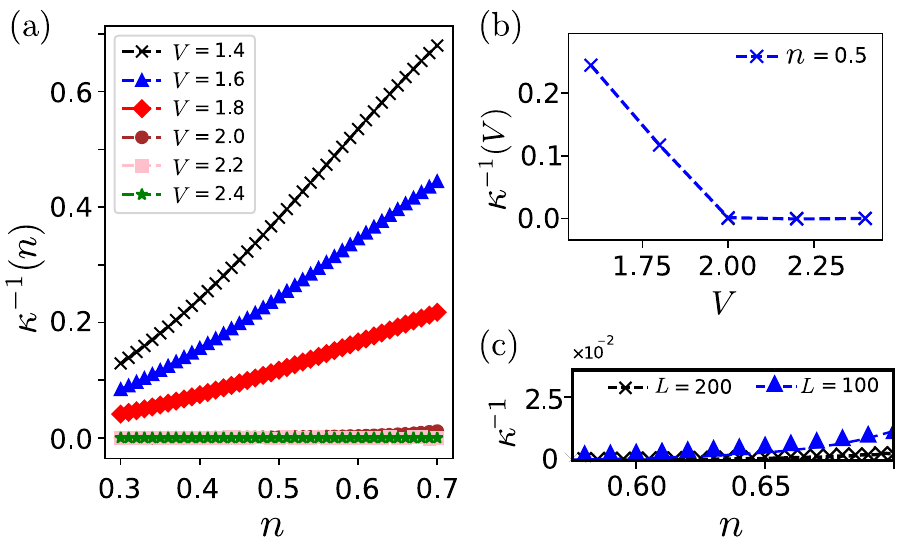}
		\caption{ The compressibility diverges for $V \ge 2$. We show compressibility inverse plotted as a function of filling $n$ and $V$, for specific $V$'s and $n$ in (a) and (b) respectively. At $V=2$, (c) depicts increasing system size improves the scaling of $\kappa^{-1}$.}\label{fig:Compr}
	\end{figure}

	We also identify another special feature of the $V=2$ point: a divergent compressibility ($\kappa$), which is the benchmark of a PS region \cite{japaninstability, Lewen2021, Moreno, Batrouni}. It is just a way of re-affirming the $K \to \infty$ limit taken on the LL theory using compressibility as a parameter \cite{GiamarchiB}. We compute $\kappa$ for a system of length $L$ with ground state energy $E_0$, where by definition:
	\begin{equation}
		\begin{aligned}\label{kapi}
			\kappa^{-1} =n^2\frac{\partial^2\mathcal{E}}{\partial n^2}= n^2 \frac{ \mathcal{E}(n + \Delta n) + \mathcal{E}(n - \Delta n) -2 \mathcal{E}(n) }{\Delta n^2}.
		\end{aligned}	
	\end{equation}
	The quantity $\mathcal{E}$ is defined as $E_0/L$. We denote  $n=\sum_j \langle n_j \rangle/L$ as the filling of the system, where $\langle \dots \rangle$ signifies a ground state expectation value. The total number of particles is $N=\sum_j n_j$. The reader may have a look at Fig.(\ref{fig:Compr}) for the results.

	We briefly turn to microscopics picture of the model to understand the origin of degeneracy as well as the divergence in $\kappa$. In $V>2$, the attractive interaction overcomes the effect of delocalization of particle density through hopping. As a result, particles prefer to bunch together, and the system goes into a phase where it is split into particle and hole like regions. The degeneracy arises because the energy of the state depends on the number of domain walls, not their position. The interaction term in $t-V$ model is diagonal in occupation representation, but the hopping is not. The number of domain walls remains fixed even if a particle is added, but the domain-length increases. In such a scenario, the modification in energy comes from the interaction term only. The change in energy becomes proportional to change in particle number and hence, the compressibility diverges.

	\section{Beyond PS}\label{sec:beyondPS}
	
	We have provided an analytic handle on the PS point of the LL phase space  hitherto inaccessible by conventional relativistic CFT methods. LL physics, closely associated to a relativistic 2d CFT, naturally goes over to a Carroll theory in the PS regime. Using symmetries of the emergent Carroll CFT, we predicted the behaviour of the density-density correlations and proceeded to verify this through an independent numerical analysis. 
	
	We now comment on what lies beyond the phase separation point. There are two ways to reach a Carroll CFT, viz. dialing the (effective) speed of light in a relativistic CFT, or by reaching a critical point in a Carroll QFT. In our free Carroll scalar theory, these correspond to $c\to0$ of massless free relativistic scalar and mass $\to 0$ of a free massive Carroll scalar. It is straightforward to see that the parameter $\epsilon$ calculated in a massive Carroll scalar theory is also equal to two. Our numerical results indicate that $\epsilon$ continues to stay at 2 beyond the PS region, deep into the other phase. 
	
\begin{figure}[ht!]
		\centering
		\includegraphics[width=.48\textwidth]{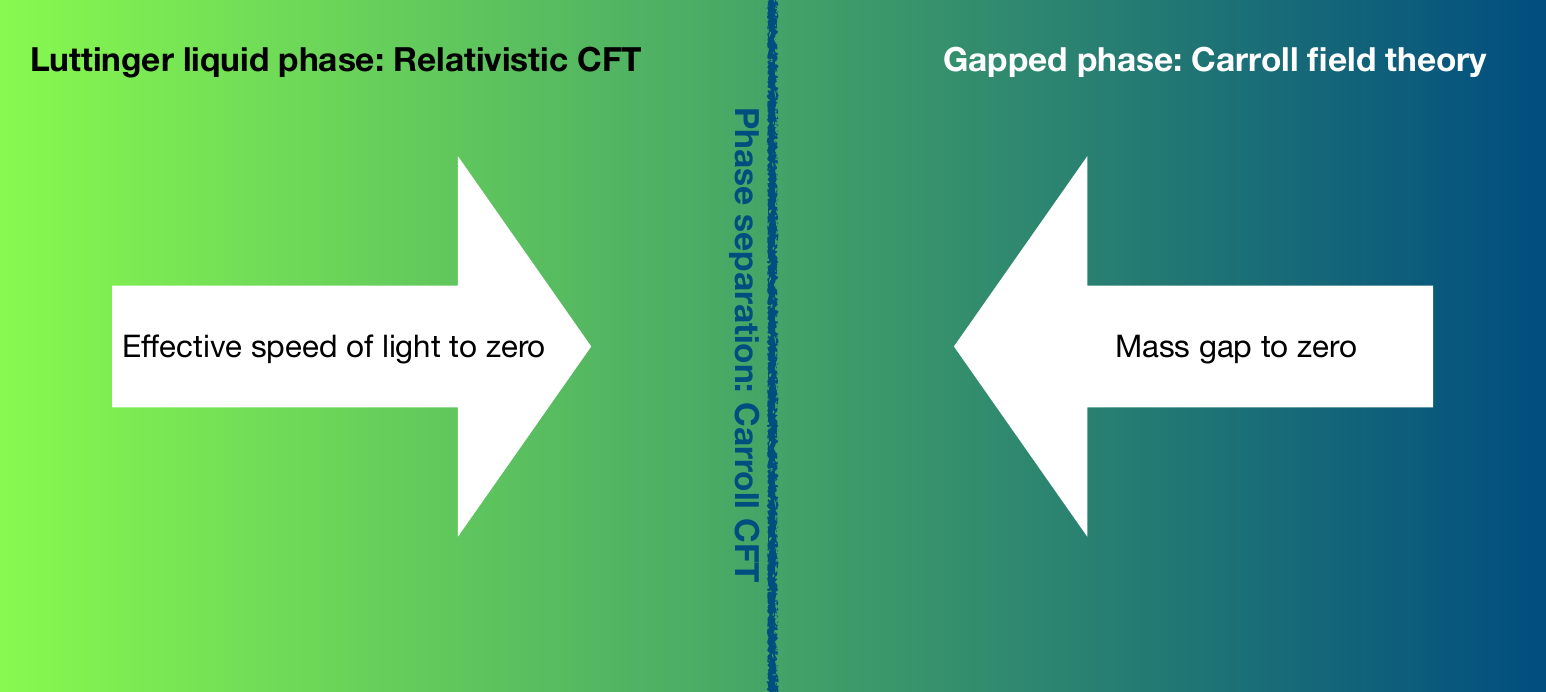}
		\caption{Approaching Carroll CFT across phase separation. }\label{PSC}
	\end{figure}

	So, it is tempting to postulate that while the LL phase is characterised by a relativistic CFT, the other phase in question is a Carroll QFT, see Fig. \eqref{PSC} for an illustration. Moving beyond the free scalar example, the Carroll phase could have a generic potential, not necessarily a mass term. The details of the system would determine the potential, but just by the virtue of this theory being Carrollian, and spatially ultralocal, we would get $\e=2$. We stress that this is a suggestive correspondence rather than an established equivalence; the gapped $V > 2$ phase also admits the standard description in terms of domain-wall configurations of an Ising ferromagnet via the XXZ mapping, and our analysis does not select between the two descriptions of the gapped phase.
	
	The emergence of a Carroll phase beyond the LL parameter space is one class of possibilities, exemplified by the model at hand. There may be gapped phases which don't correspond to a massive Carroll phase. However, irrespective of the other phase, it is clear that the PS boundary is governed by a Carroll CFT. 
	
	\section{Discussion}\label{sec:discussion}

We close by summarising what has been established directly and what is interpretive. The lattice results --- the crossover of the small-$q$ structure factor exponent from $1$ to $2$, the divergence of the compressibility, the particle--hole-segregated local density, and the extensive low-energy manifold of domain-wall configurations --- are direct outputs of the DMRG analysis, observed both in the $t$--$V$ model and in the fermionic $tJ$ model. The reduction of the bosonized Hamiltonian to that of an electric Carrollian scalar at $u \to 0$ is a continuum statement that follows directly from the form of Eq.~\eqref{masterH}. The identification of the PS boundary with an emergent Carrollian CFT is interpretive: it rests on matching the lattice observations with the continuum predictions of the Carrollian theory, and its principal added value over the conventional ``vanishing sound velocity'' picture lies in giving a single structural origin for the simultaneous appearance of the $q^{2}$ scaling, the divergent compressibility, and the supertranslation-related degeneracy. Whether all phase-separation transitions out of relativistic CFT phases are governed by Carrollian CFTs is a stronger statement that we do not establish here; we present two examples, with distinct microscopic content, in which the same scaling appears, and we view a model-independent proof of such a general claim as a question for future work.

The result of the exponent of the static structure factor, $\epsilon$, at the long wave-length limit is remarkably universal, and solely depends on the fact that the LL description terminates in a regime dictated by Carroll CFT, before entering into a (possibly) broken symmetry phase. In addition, the Carroll CFT predicts infinite ground-state degeneracy as a result of the emergent supertranslation symmetry. Similar scenario can emerge in several other physical systems, particularly in states with flat dispersion\footnote{See for example \cite{leykam2024flat} for a review and the references within.}. It has already been shown in \cite{Bagchi:2022eui,Ara:2024fbr} that Carroll symmetries arise naturally in systems with flat bands. It will be extremely interesting to explore in further detail the relevance of Carroll physics in the context of magic angle m\`{o}ire systems \cite{andrei2020graphene} or Fermi arcs in Weyl semimetals \cite{RevModPhys.90.015001}.

We have seen the emergence of Carroll physics at the edge of the LL parameter space and specifically how Carroll CFT dictates the phase separation region. The pertinent question is {\em{how are Carroll CFTs related to phase transitions in general?}} We conjecture, based on the examples studied here, that whenever there is a phase transition from a gapless phase governed by a relativistic CFT whose effective velocity is driven to zero at the transition, the phase transition point is naturally captured by a Carroll CFT. To provide non-trivial evidence for this claim, we have explored the fermionic $tJ$ model \cite{GiamarchiB, FradkinB, NagaosaQFT}, details of the study are given in the Supplementary Material \cite{Suppl}. Even though the microscopic excitations are different in nature, at the phase-separated region the correlations behave as predicted by the Carroll CFTs, supporting our claim.

It would be rewarding to look for signatures of Carroll physics in physically realizable set-ups. It requires, in case of $t-V$ model, modulation of $V$ (Fig.~\ref{fig:ExDer}, \ref{fig:Compr}(b)) in an experiment. This is analogous to tuning the coupling between $\hat{z}$-components of neighboring spins for a XXZ spin chain. In this regard, a platform offering control and tunability over the model parameters, becomes most convenient. Synthetic quantum systems have evolved over the years to qualify best for this case. Rydberg atoms \cite{PRXQuantum.3.020303, PhysRevX.8.011032} and quantum spin transistors \cite{marchu1} have shown potentials for simulating spin-1/2 XXZ chain. Other than that, ultracold atoms can also be used to study spin chains \cite{PhysRevLett.115.215301} and fermionic systems \cite{jotzu2014experimental,hirthe2023magnetically}. One can also think of dipolar system \cite{dutta2015non,chomaz2022dipolar} as a candidate for studying spin-polarised (spinless) fermions in a controlled environment. Direct experimental observations of the PS region remain scarce, as it has so far been explored only as an intermediate regime between two different phases. In this work, we tried to pursue the PS region from intrinsic symmetries, which itself is remarkably feature-rich and dictated by universal predictions, and we hope this will initiate more effort in understanding the same.

The phenomenon of emergence of Carroll CFTs at phase boundaries may have wide ranging applications even beyond condensed matter physics. We conclude with a couple of potential examples, which we mark as speculative. A leading candidate for a theory of quantum gravity is string theory. It has long been understood that if strings are heated beyond a certain limiting temperature, called the Hagedorn temperature, the conventional theory stops making sense and a new phase of string theory, called the Hagedorn phase is thought to come into existence \cite{Atick:1988si}. Due to the blowing up of the string partition function and the failure of the conventional theory, this phase has been notoriously hard to comprehend. It has recently been shown that when strings are heated to near the Hagedorn temperature, they become \textit{tensionless} \cite{tens2,tens4}. The string worldsheet is a 2D massless relativistic scalar field theory. In the tensionless limit, mirroring our current paper, a massless Carroll scalar theory arises. This suggests that the Hagedorn transition, much like the PS region of the LL theory, may be dictated by an emergent Carroll CFT. It is tempting to postulate, based on the parallel above, that the mysterious Hagedorn phase is a Carroll QFT.

Our final speculation is about Cosmology. It has been shown that in the theory of inflation, the exponential expansion of the universe in its very first fraction of a second of existence, Carroll symmetries play a role at so-called horizon crossing and super-Hubble scales \cite{deBoer:2021jej}. Cosmological phase transitions between super-Hubble and sub-Hubble scales may follow a very similar pattern to what we have discussed in our paper. These are all questions under investigation.

	\section*{Acknowledgements}
	We thank Rudranil Basu for interesting discussions and comments on the manuscript.
	
	SM is supported by grant number 09/092(1039)/2019-EMR-I from Council of Scientific and Industrial Research (CSIR).
	ABan is supported in part by an OPERA grant and a seed grant NFSG/PIL/2023/P3816 from BITS-Pilani, and further an early career research grant ANRF/ECRG/2024/002604/PMS
	from ANRF India. He also acknowledges financial support from the Asia Pacific Center for Theoretical Physics (APCTP) via an Associate Fellowship. Part of the results in this work were presented in the APCTP/GIST workshop “New Avenues in Quantum Many-body Physics and Holography” in December 2024.  AB is partially supported by a Swarnajayanti Fellowship from the Science and Engineering Research Board (SERB) under grant SB/SJF/2019-20/08 and also by an ANRF grant CRG/2022/006165.

	\bibliography{PS}
\appendix
    
	\section{Carroll Conformal algebra}\label{app:carrollalg}
	The group consists of time and space translations $(H, P_i)$, rotations $(J_{ij})$ and Carroll boosts $(C_i)$. Non-zero commutation relations are
	\begin{eqnarray}\label{carroll}
		&&[J_{ij},J_{kl}]=4\delta_{[i[k}J_{l]j]}, \, [J_{ij},P_{k}]=2\delta_{k[j}P_{i]}, \\ \nonumber &&[J_{ij},C_{k}]=2\delta_{k[j}C_{i]}, \, [C_i,P_j]=-\delta_{ij}H. 
	\end{eqnarray}
	
	Conformal Carroll group additionally contains dilatations $(D)$ and Carroll special conformal transformations $(K_0, K_i)$. Further non-vanishing commutation relations are:
	\begin{eqnarray}
		&&[D,P_i] = P_i, \, [D,H] = H ,\, [D,K_i] = -K_i, \, \nonumber \\
		&&[D,K_0] = -K_0, \, [K_0,P_i] = -2C_i, \, [K_i,H]=-2C_i, \, \nonumber \\
		&&[K_i,P_j] = -2\delta_{ij}D-2J_{ij}.
	\end{eqnarray}
	
	\section{Limiting analysis for correlators}\label{app:limits}
	As Carroll symmetry can be obtained by taking the speed of light $c\rightarrow 0$ limit from the Poincaré symmetry, one can expect to find the Carroll correlators starting from the relativistic one via the limiting procedure.  
	
	One can start with the massless relativistic scalar field theory. Then the Green's function would read
	\begin{eqnarray}
		G({x}-{x'}) = \int \frac{d^2p}{(2\pi)^2} \frac{e^{ip\cdot(x-x')}}{p^2}
	\end{eqnarray}
	Explicitly writing the $c$ factors $ G({x}-{x'}) $ becomes:
	\begin{eqnarray}
		\int \frac{dE dp}{c(2\pi)^2}\, \frac{1}{\frac{E^2}{c^2}-p^2}\exp{(i\frac{E}{c}c(t - t')-ip\cdot(x-x'))}\nn
	\end{eqnarray}
	Defining $G^{(c)}(x,t;x',t') = \frac{1}{c}G({x}-{x'})$. Now taking the Carroll limit we get
	\begin{eqnarray}\label{delta}
		G^{(c)}(x,t;x',t') &=& \int \frac{dE}{2\pi} \frac{e^{iE(t-t')}}{E^2} \int \frac{dp}{2\pi} e^{-ip(x-x')} \nn\\
		&=& -\frac{i}{2}\text{sign}(t-t') (t-t') \delta(x-x')
	\end{eqnarray}
	This can also be obtained by taking limit from a massive Carrollian theory. The two-point function of the massive theory is given by, \cite{Bagchi:2022emh,deBoer:2023fnj,Banerjee:2023jpi}
	\begin{eqnarray}
		\langle \phi(t,x) \phi(t',x')\rangle &&= \frac{1}{2m}e^{-im|t-t'|}\delta(x-x')
	\end{eqnarray}
	the small $m$ expansion of which gives $\sim \left(\frac{1}{2m}-\frac{i}{2}|t-t'| \right)\delta(x-x')$, with a divergent term in the leading order.
	Focusing on the principal value
	however gives us the more illuminating result
	\begin{equation}
		-\frac{i}{m}\sin(m(t-t'))\delta(x-x').
	\end{equation}
	Now one can take the massless limit in the above expression which gives us the finite result $\sim (t-t') \delta(x-x')$.

	\section{$t-V$ model: Ground state at $V=2$}\label{App:CarrCMT}
	As mentioned in the main text, at $V=2$ the strong attractive interactions start to take over, and particle clusters start to form.
	The ground state at $V=2$ can be thought of as a equi-probable superposition of all possible $ ({}^{L}\mathbb{C}_N)$ configurations of $N$ fermions on $L$ sites, defined at a certain filling. This state has been referred to as a ``flat state'' \cite{Casino2019}. Such a state will be flat in occupation number representation. One can write the state as a superposition
	\begin{equation}\label{GS_CarrCMT}
		\mid \mathcal{G} \rangle = ({}^{L}\mathbb{C}_N)^{-1/2} \sum_{\forall {\bf \mathcal{F}}}  \mid \mathcal{F}^L_N \rangle 
	\end{equation}
	where $\mathcal{F}^L_N$ refers to a particular sequence of possible empty and occupied sites allowed for a specific filling $N/L$ and ${\bf \mathcal{F}}$ is the set of all such possible configurations. By choosing appropriate boundary condition one can also show that energy of such a state is $-2N$ ($-2Nt$ with $t$ being set to 1). We have verified this statement from exact diagonalization computation. The crucial aspect of this result is that, change in energy due to addition of particles is proportional to number of particles added, in reference to a particular filling. It is straightforward to see that for such a situation the compressibility $\kappa$ \eqref{kapi} diverges. Hence, although $\kappa$ diverges the ground state has a separate character compared to the $V>2$ case, where the ground state is spontaneously chosen to be a distinct domain wall. 
	
	The state can be projected onto a bi-partite state where partition $p$ is of length $l$ has $m$ particles and partition $\bar{p}$ is of length $L-l$ with $N-m$ particles. In these two sectors ${}^{l}\mathbb{C}_m$ and ${}^{L-l}\mathbb{C}_{N-m}$
	configurations are possible separately, at a fixed filling of $N/L$. As a result, by inspection, one can construct a Schmidt decomposition 
	\begin{equation}\label{GS_CarrCMT_SD}
		\mid \mathcal{G} \rangle = ({}^{L}\mathbb{C}_N)^{-1/2} \sum_{m} \sqrt{ {}^{l}\mathbb{C}_m {}^{L-l}\mathbb{C}_{N-m} }  \mid \mathcal{F}^l_m \rangle_{p} \mid \mathcal{F}^{L-l}_{N-m} \rangle_{\bar{p}}.
	\end{equation}
	Here the $m$ particle state $ \mid \mathcal{F}^l_m \rangle$ and the $(N-m)$ particle state $\mid \mathcal{F}^{L-l}_{N-m} \rangle$ are separately normalized and contain equal superposition of all configurations involving the respective number of particles.
	The projected 
	reduced density matrix acquired by tracing out the partition $\bar{p}$, written as $\rho_{p} = \mid \mathcal{F}^l_m \rangle \langle \mathcal{F}^l_m \mid$, is just a pure state. Hence, Rényi entanglement entropy of order $\alpha$ written as $\mathbb{S}_{\alpha} (\rho_{p})$ is $0$, for any $m$. This can be compared to electric Carroll theories in the continuum where the ultralocality makes sure there is no coupling between neighbouring excitations, pushing the subregion entanglement entropy for ground state of such theories to an apparent zero value.
	
	\begin{figure}[htb!]
		\centering
		\includegraphics[width=.32\textwidth]{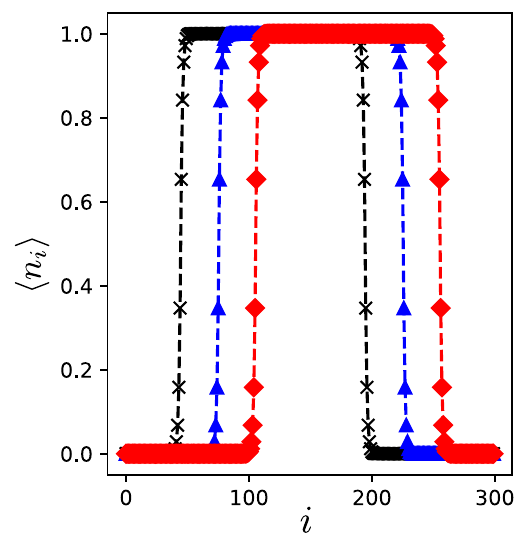}
		\caption{ For $V>$  2, the lowest energy eigenstate resides in degenerate manifold. We show the density profile for three orthogonal states from the manifold. Shifting between these states costs no energy.}\label{fig:VacDen}
	\end{figure}

	\section{Vacuum Degeneracy at $V>2$}\label{App:VacDen}
	The $V>2$ sector of our model is extensively degenerate. It is best understood in terms of an XXZ spin chain at fixed value of magnetization. This phase maps directly to the Ising ferromagnet (FM) \cite{japaninstability,GiamarchiB,Franchini2017,Cabra2004}. As discussed in the main text, for our case in the regime of interest, the lattice system behaves as if divided into particle-like and hole-like sectors separated by domain walls. This on the other hand is equivalent to clusters of up and down spins separated by the same number of domain walls in an Ising FM. It is well known \cite{Franchini2017,Cabra2004} that shifting the domain wall does not cost any energy and as a result, there are several configurations associated with the same fixed value of minimum energy. One of these configurations is spontaneously chosen as the ground state for this phase, as these translation invariant states form a subspace of the ground state Hilbert space. In Fig.~(\ref{fig:VacDen}) we show three such states from the manifold of lowest energy, picked by appropriate pinning potential. 
	
	\section{Results on the $tJ$ model}\label{App:tJMod}
	
	We bring attention to a specific interacting fermionic lattice system, namely the $tJ$ model \cite{GiamarchiB, FradkinB, NagaosaQFT}. This model is obtained by performing a strong-coupling expansion of the fermionic Hubbard model. In this model, fermions with the same spins are allowed to hop between adjacent sites with strength $t$, and neighboring localized magnetic moments are coupled via coupling strength $J$. The Hamiltonian is written as
	\begin{align}\label{H_tJ}
		\mathcal{H}_{tJ} &= -t \sum_{i,\sigma} ( f^{\dagger}_{i,\sigma} f_{i+1, \sigma} + f^{\dagger}_{i+1,\sigma} f_{i,\sigma} ) \nonumber \\&+ J \sum_i (\vec{S}_{i}.\vec{S}_{i+1} - \frac14 n_i n_{i+1}),
	\end{align}
	where $f_{i, \alpha}$ is an annihilation operator associated with creating a fermion with spin $\alpha (=\uparrow, \downarrow)$ at site $i$. The operator $n_i = \sum_{\alpha} f^{\dagger}_{i, \alpha}  f_{i, \alpha}$ denotes local density of fermions. The local moments are defined by the spin operator $\vec{S}_i = f^{\dagger}_{i, \alpha} \vec{\sigma}^{\alpha \beta} f_{i, \beta}$, with Pauli matrices $\vec{\sigma}$. In what follows, we set $t=1$ for all the numerical computations.
	
	The phase diagram of the $tJ$ model is known to host a phase where the compressibility is divergent and the local density is distinctly divided into particle and hole-like regions. This has been referred to as the electron-solid phase-separated (PS(ES)) region \cite{Man2011}, and can be obtained by tuning $J$ starting from a phase governed by a relativistic CFT. There are several similarities among the properties of the phases related by this transition, to the ones found in our study of the $t-V$ model. As a result, we are compelled to investigate further into this system, in spite of microscopic differences.

	\begin{figure}[htb!]
		\centering
		\includegraphics[width=.425\textwidth]{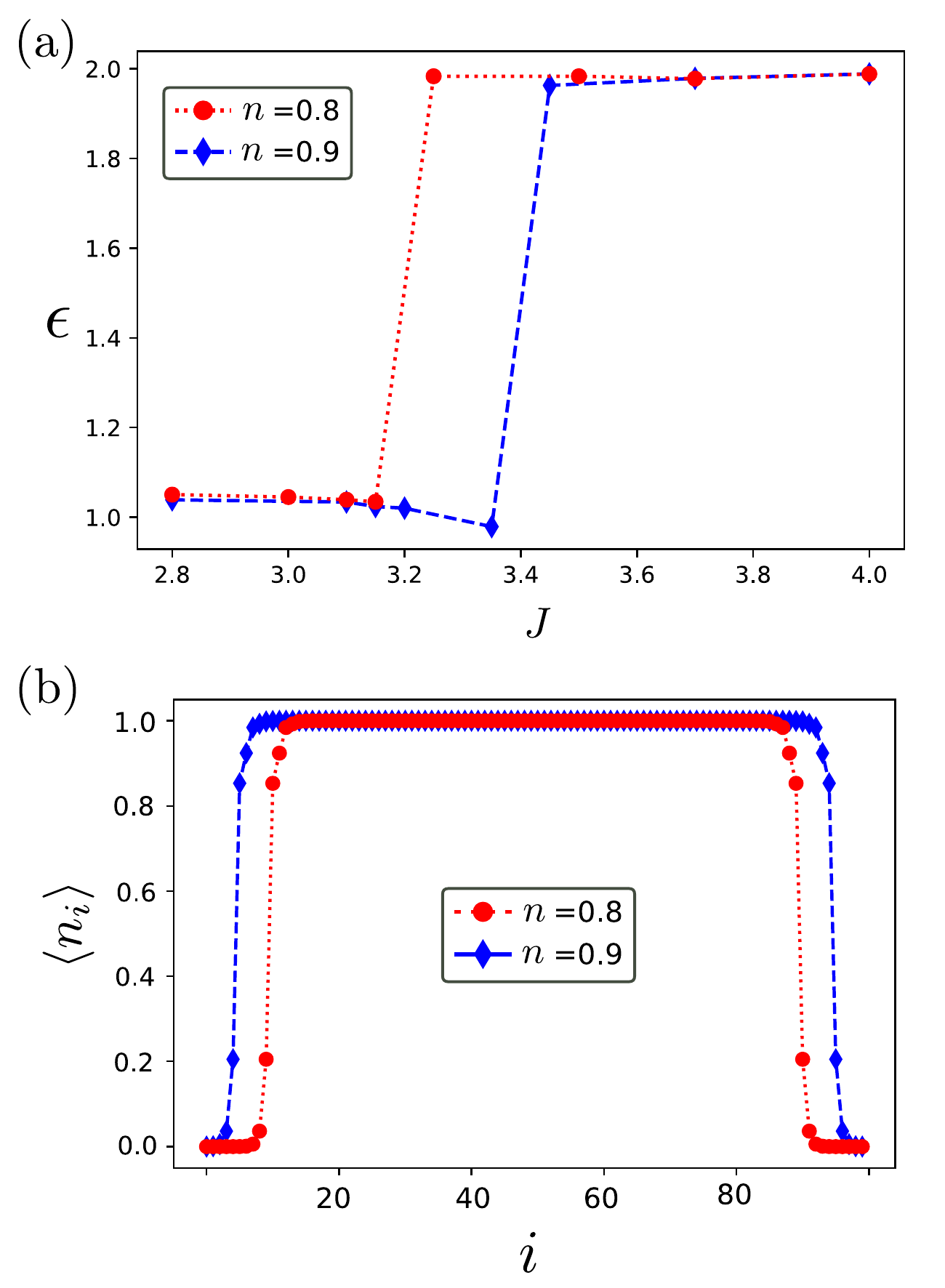}
		\caption{ The principal features of the phase transition discussed using the $t-V$ model also show up in the fermionic $tJ$ model. In (a), we see a discontinuous transition between a gapless Luttinger liquid-like phase and a phase-separated electron-solid (PS(ES)) region.  In (b), the local particle density is shown divided between particle and hole-like regions, similar to Fig.~\eqref{fig:VacDen}. We emphasize that the exponents always remain close to $1$ and $2$. The small deviations are numerical artifacts, and we don't go into their details as we have already presented an instance of finite size scaling for $t-V$ model in Fig.~(\ref{fig:ExDer}).}\label{fig:tJ}
	\end{figure}
	
	We study the model by scanning across $J$ for comparatively higher filling ($i.e.$ $n=0.8,0.9$) of the system. We choose to concentrate on the particular region relevant to our study. The phase transition under discussion is depicted in Fig.~(\ref{fig:tJ}). We observe a discontinuous transition in terms of the scaling exponent $\epsilon$. The $\epsilon=1$ segment is governed by the laws of a gapless Luttinger liquid or the relativistic CFT. The $\epsilon=2$ region pertains to PS(ES). The local density profile in the PS(ES) is also shown. The transition point between these two phases is dictated by Carrollian CFT, which envisages the correct scaling behavior of this transition. It implies that the transitions shown in Fig.~(\ref{fig:ExDer}) and Fig.~(\ref{fig:tJ}) both carry signatures of the underlying Carrollian physics despite emanating from completely different lattice systems. 
\end{document}